\documentclass[final,5p,times]{elsarticle}

\usepackage{amssymb}
\usepackage{amsmath}
\usepackage[greek.ancient,ngerman,english]{babel}
\usepackage{siunitx}
\DeclareSIUnit\inch{"}
\DeclareSIUnit \ADCchannel {ADC\ channel}
\usepackage{hyperref}
\usepackage{cleveref}
\usepackage[dvipsnames]{xcolor}
\usepackage{textgreek}
\usepackage[autostyle, english = british]{csquotes}
\usepackage{booktabs}
\usepackage{multirow}
\usepackage{multicol}
\usepackage{subcaption}

\newcommand{\isotope}[2]{\textsuperscript{#1}{#2}}

\graphicspath{{pictures/}}
\journal{Nuclear Instruments and Methods Section A}

\begin{document}

\begin{frontmatter}

\title{Characterisation of a Thick Pixelated Silicon Detector for Electron Spectroscopy of Neutron Beta Decay}

\author[frm,tum1]{Manuel Lebert}
\author[tum1]{Igor Konorov}
\author[tum1]{Lilli Löbell}
\author[tum1]{Bastian Märkisch}
\ead{maerkisch@ph.tum.de}

\affiliation[frm]{organization={Heinz Maier-Leibnitz Zentrum (MLZ), Technical University of Munich},
            addressline={Lichtenbergstr.~1}, 
            city={Garching},
            postcode={85748},
            country={Germany}}

\affiliation[tum1]{organization={Technical University of Munich, School of Natural Sciences},
            addressline={James-Franck-Str.~1}, 
            city={Garching},
            postcode={85748},
            country={Germany}}

\begin{abstract}
    Silicon detectors are commonly used for spectroscopy of low-energy particles. For electrons in the \SI{1}{\MeV} range, a rather large thickness of $\approx\qty{2}{\milli\meter}$ is required to entirely stop the electrons and commercial options are scarce.
    With the instrument PERC at the FRM~II, we aim to measure beta spectra from polarised and unpolarised neutrons in order to determine the axial-vector coupling constant, the element $V_\textrm{ud}$ of the Cabibbo-Kobayashi-Maskawa quark-mixing matrix, and to search for hypothetical scalar and tensor contributions. 
    We present the characterisation of a commercially available, pixelated detector to assess its suitability to measure the entire electron energy spectrum of free neutron beta decay.
\end{abstract}




\end{frontmatter}

\makeatletter
\def\ps@pprintTitle{%
  \let\@oddhead\@empty
  \let\@evenhead\@empty
  \def\@oddfoot{\reset@font\hfil\thepage\hfil}
  \let\@evenfoot\@oddfoot
}
\makeatother

\section{Introduction}
    \label{sec:introduction}

    The PERC facility \cite{Dubbers.2008} at the MEPHISTO beam line \cite{Klenke.2015} of the MLZ, Garching, will be used to investigate the decay of free neutrons via measurements of correlation parameters \cite{Jackson.1957}. The main objectives are to measure the parity-violating beta asymmetry parameter $A$ and the Fierz interference term $b$. PERC aims at a five-fold improvement on the precision of the beta asymmetry over the most precise result by \textsc{Perkeo~III} \cite{Markisch.2019} to determine the element $V_\mathrm{ud}$ of the Cabibbo-Kobayashi-Maskawa matrix to address the current Cabibbo-angle anomaly \cite{Seng.2018,Grossman.2020}, as well to improve current best limits on the Fierz interference term $b$ \cite{Saul.2020, Beck.2024} to a precision of $\Delta b = 1 \times 10^{-3}$. 
    
    At PERC, neutron decay is observed in an \qty{8}{\meter} long section of the \qty{12}{\meter} long magnet system \cite{Wang.2019} in a magnetic field of typically about \qty{1.5}{T}. 
    Electrons and protons emitted within a certain solid angle are guided by the magnetic field towards the main detection system at the downstream end of the magnet system. A magnetic filter between the decay region and the main detection system with a field strength of up to \qty{6}{T} allows only about \qty{7}{\percent} of the electrons to get to the main detector. Behind the filter, the magnetic field drops to typically \qty{0.375}{T} where the detector system is placed. The silicon detector characterised in this work is considered as a future upgrade to replace a plastic scintillator based detector due to its much better energy resolution and linearity.

    On the upstream side of the decay volume, a split detection system serves to absorb the major part of the remaining particles and to identify electrons backscattered from the main detector system. This upstream detector consists of two plastic scintillators read out from the back side by SiPM arrays enabling spatial resolution \cite{Ziener.2014, Bernert.2025}. The identification of backscatter events can be much improved if also the main detector offers a spatial resolution of $\approx$\,\qty{1}{\centi\meter}. It requires a time resolution of $\mathcal{O}(\qty{10}{\nano\second})$ for both detectors in order to use time-of-flight to correlate events on the up- and downstream detectors. Time-of-flight can also be used to calibrate the detectors \cite{Roick.2018,Bernert.2025}.
    
    The downstream detector will initially be a scintillation detector, which is read out from the back by \num{7} PMTs. An upgrade to a pixelated silicon PIN-detector has the advantage of a much better linearity (\cref{sec:photon_calibration}) \cite{Zulliger.1969} as quenching effects common to organic scintillators \cite{Birks.1951} are absent. Silicon detectors also have a much better energy resolution (see \cref{sec:photon_calibration,sec:electrondetection}). Plastic scintillators on the other hand show a much lower probability to backscatter electrons, provide superior timing, and are less susceptible to gamma background radiation.  For PERC, timing requirements are much relaxed as compared to its predecessors \textsc{Perkeo~II} \cite{Abele.1997} and \textsc{Perkeo~III} \cite{Maerkisch.2009}, and sensitivity to background is reduced by the design of PERC \cite{Dubbers.2008,Wang.2019}. A reliable characterisation of the energy-response of the detector on the level of $10^{-4}$ is required for a precision of $\Delta b = 1\times 10^{-3}$.
    
    The maximum kinetic energy of the electrons from free neutron decay is \qty{782}{\keV}.
    The energy of a conversion electron peak of \isotope{207}{Bi},  which will be used for calibration, is \qty{975.651}{\keV} \cite{Kondev.2011}.
    In the continuous-slowing-down approximation (CSDA), these electrons have a range in silicon of $\approx$\,\qty{1.7}{\milli\meter} and  $\approx$\,\qty{2.3}{\milli\meter}, respectively \cite{Seltzer.1993}.
    Because of scattering inside the detector, it does not need to be as thick as the CSDA range, but according to our simulations using PENELOPE \cite{Penelope.2019}, it needs to be \qty{2}{\milli\meter} thick.
    
    The Nab collaboration aims to measure the electron-neutrino correlation parameter $a$ and the Fierz interference term $b$ at the Spallation Neutron Source, Oak Ridge, Tennessee \cite{Baessler.2024}. In a follow-up measurement with polarised neutrons, the instrument will be used to determine the beta asymmetry parameter \cite{Baessler.2025}.  The Nab, UCNB \cite{Broussard.2017}, PANDA, and abBA \cite{Wilburn.2005} collaborations jointly developed a pixelated thick silicon detector in collaboration with Micron Semiconductor Ltd.\footnote{\url{https://www.micronsemiconductor.co.uk/}}, which is commercially available as model MSPX128. Its performance with an emphasis on the detection of low-energy ions, protons-electron coincidence, and timing has been investigated \cite{SalasBacci.2014, Broussard.2017, Hayen.2023, Taylor.2025}. In this paper, we focus on the calibration with photons and the detection of electrons up to an energy of about \qty{1}{MeV}.
    
\section{Detector Design}
    \label{sec:detector_design}

    \begin{figure}
        \centering
        \includegraphics[width=\linewidth]{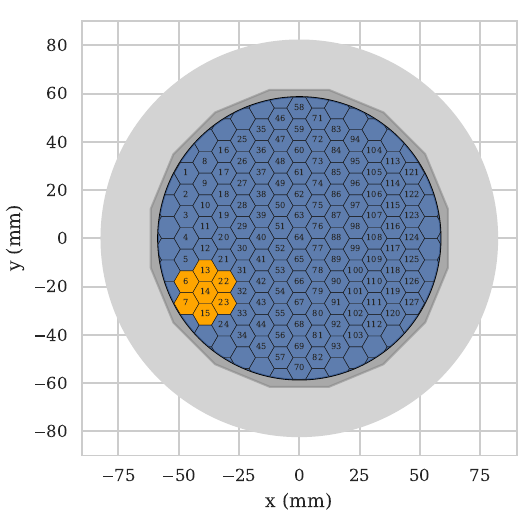}
        \caption{Depiction of the anode side of the silicon detector.
        The \num{127} pixels are numbered.
        The black circle around the pixels is the end of the active area and the position of the guard ring.
        The incomplete pixels are connected to a single pin for biasing and cannot be read out.
        The gray polygon is the silicon wafer, and the white disc is the ceramic backing.
        The pixels highlighted in orange are analysed in this work.}
        \label{fig:silicon_detector}
    \end{figure}
    The pixelated detector is a polygonal cut from a \qty{6}{\inch} silicon wafer, which is glued to a \qty{1}{\milli\meter} thick ceramic disc.
    The active area has a diameter of \qty{117.15}{\milli\meter}.
    The entrance side of the detector is continuously boron-doped with a nominal depth of \qty{100}{\nano\meter}.
    An aluminium grid for electrical connection is sputtered on the entrance window.
    It covers roughly \qty{1}{\percent} of the area and is \qty{300}{\nano\meter} thick.
    The back side has a region doped with phosphor of \qty{300}{\nm} thickness and is pixelated. 
    Each pixel is hexagon-shaped with a maximal diameter of \qty{10.265}{\mm} and an area of \qty{70}{\mm\squared}.
    A depiction of the pixel layout is shown in \cref{fig:silicon_detector}.
    The pixels are separated by a distance of \qty{100}{\um}. 
    Due to the continuous p\textsuperscript{+} layer at the entrance window, no dead region is present.
    The electric field lines extend from the p--n junction to the anode, thereby covering the entire active volume of the device.
    Near the pixel boundaries, the field lines split and terminate on different pixel anodes.
    As a consequence, charge-sharing occurs when free charge carriers are generated but are not collected exclusively by a single anode.
    However, the free charge carriers are not trapped within the detector volume~\cite{Hayen.2023}. 
    The detector is available with pin-header connectors or with connection points for spring-loaded pins on the back side of the ceramics. In our case, four pin headers are used to bias and read out the pixels.
    There are \num{127} complete pixels, which can be read out separately. 
    All incomplete pixels at the edge of the active area are connected together to a single pin, allowing them to be biased but not read out. 
    Around the active area of the detector is a guard ring, which protects the active area from charge carriers in the non-active volume of the silicon.

\section{Measurement Setups}

    In order to characterise the detector, three different configurations were used, which are described below: the \enquote{leakage current setup}, the \enquote{calibration setup}, and the \enquote{dead layer setup}.
    All measurements were performed at room temperature in a temperature controlled environment.
    For all configurations, the bias voltages are supplied by a single Mesytec\footnote{\url{https://www.mesytec.com/}} MVHV-4 module with four channels.
    To measure the detector's output, a \num{16} channel Mesytec MPR-16 preamplifier and an eight-channel multichannel analyser CAEN\footnote{\url{https://www.caen.it/}} V1782 were used and read out with the manufacturer's software CoMPASS. The setup so far lacks an appropriate low-pass filter which in combination with the MCA's RC‐CR\textsuperscript{2} filter leads to event triggers on high-frequency noise. Unfortunately, we also observed noise in the frequency range of the signal.
    
    \paragraph{Leakage current setup}
        This was used to measure the leakage current of the pixels and the guard ring.
        The detector was contained in an electrically grounded metallic box at atmospheric pressure to shield it from light and electromagnetic interference.
        The detector's four pin headers are each connected to an adapter PCB.  Each PCB has a connection for the bias-voltage supply, which supplies each pixel via a \qty{10}{\mega\ohm} resistor. To determine the leakage current, the voltage drop across these resistors was measured and the leakage current was calculated using Ohm's law.
        The guard ring is biased via a separate channel of the power-supply and  its leakage current can be read out directly.
        
    \paragraph{Calibration setup}
        This was used for the energy calibration using photons and the charge-sharing measurements. 
        The detector is contained in the same metallic box at atmospheric pressure as for the \enquote{leakage current setup}. The active side is facing downward.
        Three of the four headers on the detector are connected via the same adapter PCBs used for the \enquote{leakage current setup}. 
        The remaining header is connected to a different PCB, with \num{16} channels connected to the preamplifiers via a flat ribbon cable and an IDC connector, while the other \num{16} pixels of this header are biased via the guard ring connection of the amplifier and \qty{10}{\mega\ohm} resistors in parallel.
        This allows to read out a maximum of \num{16} pixels simultaneously. Since the DAQ currently only has eight channels,  we only read out one central channel and its six direct neighbours as can be seen in \cref{fig:silicon_detector}. The selection of pixels includes both pixels close to the guard ring and pixels more toward the centre of the detector. 
        Beneath the detector, different radioactive sources can be placed at different distances. Typically, a distance between the source and the detector of \qty{1}{\cm} was used.
        
    \paragraph{Dead layer setup}
        This setup was used to determine the effective thickness of the dead layer using a radioactive alpha source.
        The detector was placed in a vacuum chamber with a volume of roughly \qty{1}{\m\cubed} at a pressure of \qty{4.5e-4}{mbar}.
        The position of the radioactive source below the detector can be adjusted using a push-pull positioner without breaking the vacuum. The same electronics and detector channels were used as for the \enquote{calibration setup}.

\section{Characterisation}

    \subsection{Leakage Current}
        \label{sec:leakage_current}

        \begin{figure}
            \centering
            \includegraphics[width=\linewidth]{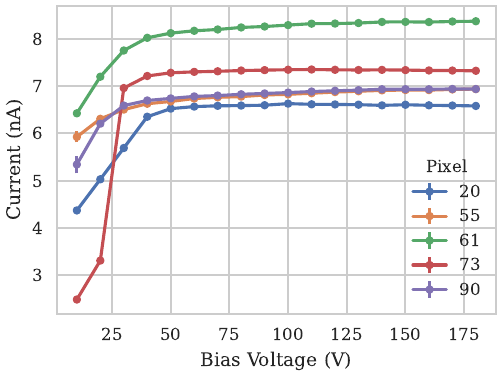}
            \caption{The leakage currents for a selection of inner pixels show the expected behaviour. Above a bias voltage of around \qty{50}{\volt}, the pixels seem fully saturated.}
            \label{fig:leakagecurrent-inner}
        \end{figure}
        \begin{figure}
            \centering
            \includegraphics[width=\linewidth]{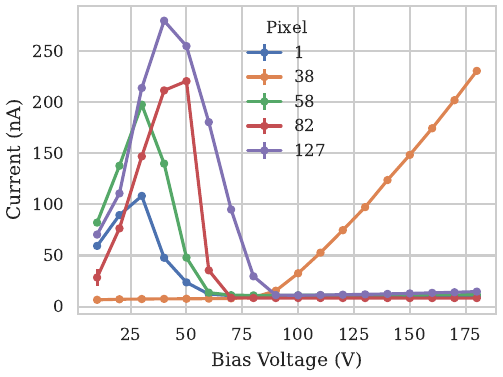}
            \caption{The leakage current of the outer pixels shows a maximum at low bias voltages. It saturates at the same level as the inner pixels for higher voltages, though. A single inner pixel exhibits ohmic behaviour.}
            \label{fig:leakagecurrent-outer}
        \end{figure}
        To measure the leakage current, the \enquote{leakage current setup} together with a Keithley\footnote{\url{https://www.tek.com/}} 3700 multimeter with an input resistance above \qty{10}{\giga\ohm} was used.
        For every value of the bias voltage, we changed the voltage and then waited for \qty{30}{\min} to allow the detector to acclimate before measuring the voltage drop across the \qty{10}{\mega\ohm} resistors 10 times.
        The leakage currents of all pixels and the guard ring are determined separately and averaged over all measurements.
        
        In a reverse-biased PIN diode, the current typically increases with the voltage until it reaches a plateau. Beyond that, the current hardly changes until the breakthrough voltage is reached \cite{Spieler.2012}. This behaviour can be observed for all but one pixel of the detector which are not adjacent to the guard ring, see \cref{fig:leakagecurrent-inner}. The leakage currents are consistently below \qty{20}{\nano\ampere}.
        
        The pixels at the edge of the detector behave differently and show a spike at low voltages, see \cref{fig:leakagecurrent-outer}. This refers not only to the incomplete pixels combined to form pixel \num{128}, but also to all pixels bordering those incomplete pixels and to the few pixels directly at the guard ring. Almost all of these pixels show a leakage current ranging from \qtyrange{50}{350}{\nano\ampere} for voltages below \qty{60}{\volt}.
        
        A single inner pixel shows an about constant leakage current up to \qty{90}{\volt} bias. For larger voltages, the current for this pixel increases linearly, corresponding to a resistance of \qty{420}{\mega\ohm}, see pixel~38 in \cref{fig:leakagecurrent-outer}. 
        This behaviour cannot be due to reaching the breakdown voltage, as one would expect first an exponential and then a linear increase, where the slope would correspond to the pixel's current-limiting resistor of only \qty{10}{\mega\ohm}.

        The leakage current of the guard ring (not shown)
         shows a nearly linear increase which might potentially be due to imperfect implantation of one of the doping atoms. This behaviour was not observed by the manufacturer, though.

        On average, the full depletion voltage of the pixels is between \qtylist{50; 60}{\volt}.
        The nominal resistivity of the detector is \qty{25}{\kilo\ohm\cm}, which would result in a depletion voltage of $\approx$\,\qty{300}{\volt}, according to the manufacturer.
        Given this significant difference in depletion voltage, it appears that the detector's doping also differs quite a bit from the nominal value, see \cref{sec:risetime}.

        A bias voltage of \qty{100}{\volt} was used for most of the subsequent characterisation steps. The maximum bias voltage was limited in order to not exceed the maximum rated current for the guard ring and to limit the signal noise.
        
    \subsection{Stability}
    \label{sec:stability}

        \begin{figure}
            \centering
            \includegraphics[width=\linewidth]{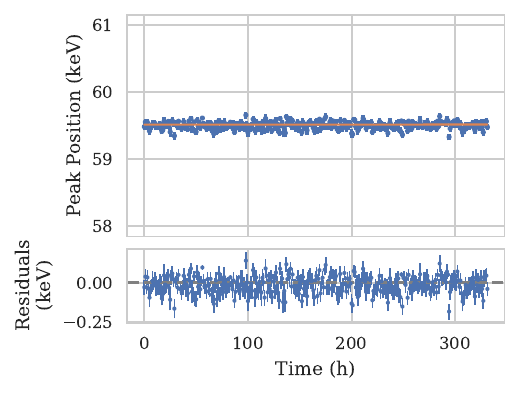}
            \caption{Position of the \qty{59.5}{\keV} photon line of \isotope{241}{Am} measured for over \qty{300}{\hour}. The orange line is a linear fit to the extracted peak positions. The drift is below \qty{0.15}{\eV \per \hour} for all pixels of the detector and \qty{0.008(24)}{\eV\per\hour} for this specific one.}
            \label{fig:stability}
        \end{figure}

        To determine the stability of the whole system, the \textquote{calibration setup} was used.
        An \isotope{241}{Am} source was enclosed in plastic, which blocked its \textalpha-radiation. The position of the photon peak at around \qty{59.5409(1)}{\keV} was measured continuously by the detector for over \qty{300}{\hour}. The data was then aggregated into chunks of \qty{1}{\hour} to determine the drift of the system.
        The peak position was determined by fitting the following function to the data:
        \begin{equation}
        \begin{aligned}
            f(x)=& \frac{A}{\sqrt{2\pi\sigma^2}} \cdot \exp{\left(-\frac{(x-\mu)^2}{2\sigma^2} \right)} \\
                &+ A_\mathrm{iec} \cdot \mathrm{erfc}\left( \frac{(x-\mu)}{\sqrt{2}\cdot \sigma} \right) \\
                & + \frac{A_b}{\sqrt{2\pi\sigma_b^2}} \cdot \exp{\left(-\frac{(x-\mu_b)^2}{2\sigma_b^2} \right)} \\
                &+ \mathrm{const.}
            \label{eq:photonpeak}
        \end{aligned}
        \end{equation}
        The first term describes the photopeak statistics. The cumulative error function of the second term accounts for incomplete energy collection, for example due to photons escaping the detector or charge carriers being trapped \cite{Routti.1969,Phillips.1976,Campbell.1985,Knoll.2010}.
        The second Gaussian distribution accounts for the backscatter peak due to Compton scattering at 180\textdegree\ in the backing material of the source at about \qty{43}{\keV}.
        The last term describes a constant background.
        
        The result for one pixel is shown in \cref{fig:stability}.
        By applying a linear fit to the time-dependence of the peak position, we observe that the drift is less than \qty{0.15}{\eV\per\hour} for all measured pixels, with most pixels at a level of \qty{0.05}{\eV\per\hour}. This is better by an order of magnitude compared to what was achieved in the latest campaign by \textsc{Perkeo~III} in 2020 with a scintillator-based detector \cite{Lamparth.2022}. Ref.~\cite{Taylor.2025} also reports no significant proton peak drift over 126 days.
        
    \subsection{Noise Characterisation and the Filter Rise-Time}

        \begin{figure}
            \centering
            \includegraphics[width=\linewidth]{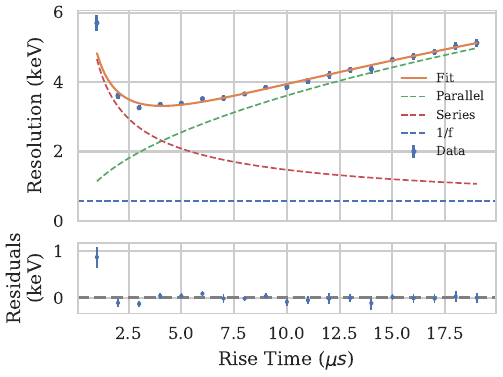}
            \caption{Energy resolution of the \qty{59.5}{\keV} photon line of \isotope{241}{Am} measured for varying rise times of the trapezoidal filter for pixel~23. Parallel and series noise are dominant at short and long rise times, respectively. The $1/f$ noise does not contribute significantly to the overall energy resolution.}
            \label{fig:resolution}
        \end{figure}

        The height of the measured signal is determined by convoluting the incoming signal a with a trapezoidal filter. The time during which the filter's output rises linearly is called the \enquote{rise time}. The height above the baseline during the \emph{flat top phase} is taken as the signal height.
        The energy resolution of the detector system is limited by statistical fluctuations in charge generation and electronic noise.
        The latter is typically described by three contributions: parallel, series, and $1/f$ noise \cite{Spieler.2012}.
        Parallel noise is associated with fluctuating input currents, like the leakage current of the silicon detector or bias-resistor noise.
        Series noise stems from voltage fluctuations at the preamplifier input and resistances in series with the detector. 
        The $1/f$ noise comes from trapping and detrapping of charge carriers and the front-end electronics of the preamplifier.
        As these three sources have different frequency spectra, their contributions to the energy resolution can be influenced by varying the rise time of the trapezoidal filter: shorter times increase the contribution of high-frequency series noise, and longer times increase the contribution of parallel noise.
        The $1/f$ noise contribution is flat, i.e., changes in the rise time do not affect its contribution \cite{Gatti.1990}.

        To determine the noise contributions and the optimal rise time, the same setup and source as in \cref{sec:stability} was used, while varying the rise time.
        We performed fits using \cref{eq:photonpeak} to the photopeak at \qty{59.5409(1)}{\keV} for the different measurements to extract the mean position $\mu$ and width $\sigma$.
        The width is a measure of the energy resolution, and its dependence on the rise time $t$ can be described as \cite{Spieler.2012}:
        \begin{equation}
        \sigma(t) = \sqrt{a\cdot t + \frac{b}{t} + c},
        \end{equation}
        where the three terms represent the parallel $a$, series $b$, and $1/f$ noise contributions $c$.
        As the peak position also depends on the rise time, it is necessary to do a rough calibration for each setting assuming a linear calibration function through the origin.
        
        \Cref{fig:resolution} shows the measured energy resolution (peak width) versus the rise time of the filter for pixel~23.
        As expected, the series and parallel noise dominate the energy resolution, while the $1/f$ contribution is nearly negligible.
        The optimal rise time is at the minimum of the fit function, i.e., where the contributions by the parallel and series noise cross.
        This is at around \qty{4}{\us}. Across all seven pixels, the smallest width was observed at \qty{3}{\us} and, thus, chosen for all subsequent measurements.
        This discrepancy is most likely due to insufficient data below \qty{1}{\us}.

        \begin{figure*}
            \begin{subfigure}{.5\textwidth}
                \centering
                \includegraphics[width=\linewidth]{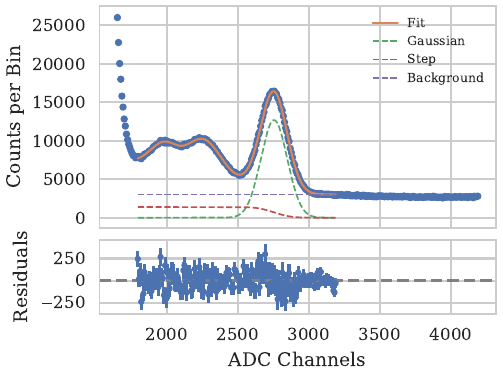}
                \caption{\qty{81}{keV} lines of \isotope{133}{Ba}}
            \end{subfigure}
            \begin{subfigure}{.5\textwidth}
                \centering
                \includegraphics[width=\linewidth]{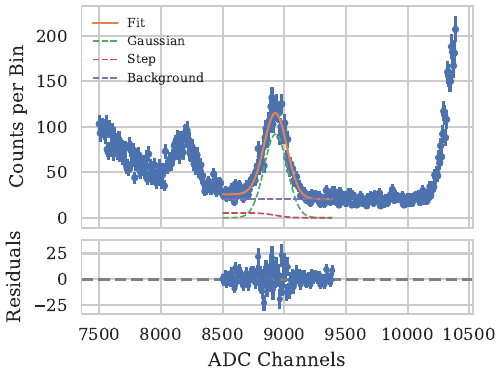}
                \caption{\qty{303}{keV} line of \isotope{133}{Ba}}
            \end{subfigure}\\
            \begin{subfigure}{.5\textwidth}
                \centering
                \includegraphics[width=\linewidth]{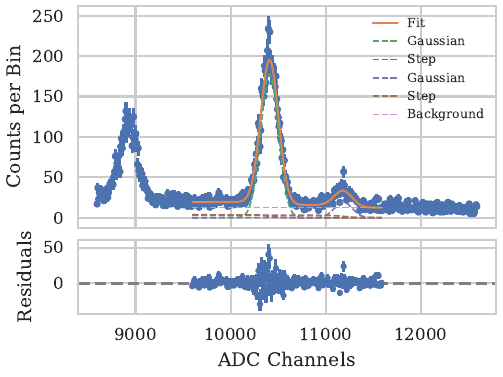}
                \caption{\qty{356}{keV} and \qty{384}{keV} lines of \isotope{133}{Ba}}
            \end{subfigure}
            \begin{subfigure}{.5\textwidth}
                \centering
                \includegraphics[width=\linewidth]{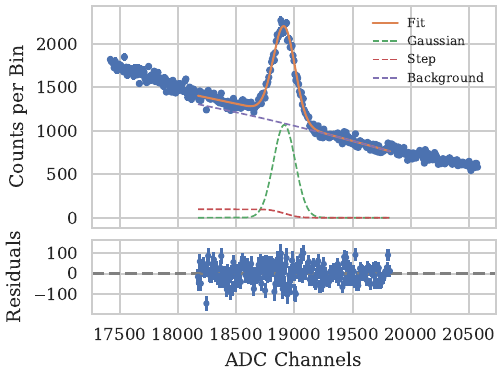}
                \caption{\qty{661}{keV} line of \isotope{137}{Cs}}
            \end{subfigure}
            \caption{Simultaneous calibration fit to the photon lines listed in \cref{tab:calpeaks}. The coloured lines indicate the fit regions. The fit of the \qty{81}{keV} lines includes the backscatter peak and another another line at about \qty{51}{keV}. The fit function for the \qty{356}{keV} and \qty{384}{keV} lines shares a common background. The background model for the \qty{661}{keV} includes a quadratic term. Figure from \cite{Lebert.2026}.}
            \label{fig:photoncalibration}
        \end{figure*}

        A way to improve the energy resolution is to reduce series noise.
        Lowering the temperature of the detector reduces the leakage current and is implemented for Nab \cite{Broussard.2017,Taylor.2025}.
        For the PERC experiment, we plan to stabilise the detector to a temperature down to \qty{-50}{\degreeCelsius} which is expected to reduce the leakage current by an order of magnitude. The energy resolution achieved here with commercial electronics at room temperature is comparable to that obtained in \cite{Taylor.2025} where the detector was operated with a bias voltage of \qty{300}{V} at cryogenic temperatures.

    \subsection{Calibration using Photons}
        \label{sec:photon_calibration}

        \begin{table}
    	    \caption[Photon Peaks]{Monoenergetic photon peaks used for the energy calibration of the silicon detector \cite{Be.2016, Mougeot.2025}.} 
    	      \centering
    	      \sisetup{table-number-alignment=center, 
    			table-figures-decimal=3,
    			add-decimal-zero
    			}
    		\begin{tabular}{cS}
    			\toprule
    			Isotope &  {Energy (\unit{\keV})} \\
    			\midrule
    			\midrule
                \multirow{5}[0]{*}{\isotope{133}{Ba}}   & 79.6142(19)\\
                                                        & 80.9979(11)\\
                                                        & 302.8512(16) \\
                                                        & 356.0134(17)\\
                                                        & 383.8491(12)\\
                \midrule
                \isotope{137}{Cs}  & 661.657(3) \\
    			\bottomrule
    		\end{tabular}
            \label{tab:calpeaks}%
        \end{table}%
        
        For detector calibration, the \textquote{calibration setup} was used.
        As radioactive sources, \isotope{133}{Ba} and \isotope{137}{Cs} were used and their photon energies are listed in \cref{tab:calpeaks}. Since the detector cannot resolve the two lines of \isotope{133}{Ba} at around \qty{80}{\keV}, these were treated as a single line with an energy corresponding to the weighted mean. 
        The strongly decreasing interaction probability via the photoeffect for photons at high energies implies that either very long measurement times or sources with high activity are required. Due to the lack of an adequate source, we could not cover the full range up to \qty{1}{MeV}.
        
        To describe the peaks in the spectrum, we use the function in \cref{eq:photonpeak}, excluding the Gaussian for the backscatter peak for all energies except for the \qty{80}{\keV} peak. Only in this case, the backscatter peak is close at around \qty{61.5}{\keV} and needs to be taken into account. Due to its proximity, it shifts the determined mean of the main peak to lower energies. This shift in the mean energy was determined using Monte-Carlo simulations to be \qty{34.7}{\eV} and included as a fixed shift in the fit function. Additionally, we include another line at \qty{53.1622 (18)}{\keV} in the fit, but do not directly use it or the backscatter peak position for the determination of the energy-channel relation. We note that the width of the backscatter peak is a free parameter.

        For all lines, we assume a different constant background. This is due to Compton spectra from other lines as well as due to the plastic surrounding the sources. 
        \isotope{133}{Ba} and \isotope{137}{Cs} also emit electrons, which are stopped in the plastic housing and produce Bremsstrahlung.
        For the spectrum from the \isotope{137}{Cs} source, we extend the model for the background with a quadratic term. 

        The width of the photon peaks $\sigma$ can be described as a combination of several components \cite{Knoll.2010}:
        \begin{equation}
            \sigma^2(E) = \sigma_\mathrm{stat}(E)^2 + \sigma_\mathrm{electronic}^2 + \sigma_\mathrm{other}^2
        \end{equation}
        $\sigma(E)_\mathrm{stat}$ describes the statistical fluctuation of the electron-hole pair creation as a function of the energy.
        It can be calculated via $\sigma_\mathrm{stat}(E)^2 = FE\epsilon$, with the Fano factor $F=\num{0.12}$ and the excitation energy per electron-hole pair in silicon $\epsilon = \qty{3.64}{\eV}$ \cite{Knoll.2010}. At an energy of \qty{1}{\MeV}, this results in $\sigma_\mathrm{stat}(\qty{1}{\MeV})^2 = \qty{0.44}{\keV\squared}$, which is very small compared to the other terms.
        The last term combines effects such as trapping or incomplete charge collection, which are insignificant here. It is hence not explicitly included in the fit.
        The dominant component of the resolution is the contribution from the electronics (e.g., noise) $\sigma_\mathrm{electronic}$, which is assumed to be independent of the energy.

        The calibration function was assumed to be linear as silicon detector systems are known for their highly linear behaviour. The fit function for the calibration has a total of \num{25} free parameters to describe all peaks, backgrounds, and a single backscatter peak simultaneously. The parameters include a common gain and offset for the linear calibration model, as well as a common electronic width. 

        \begin{figure}
            \centering
            \includegraphics[width=\linewidth]{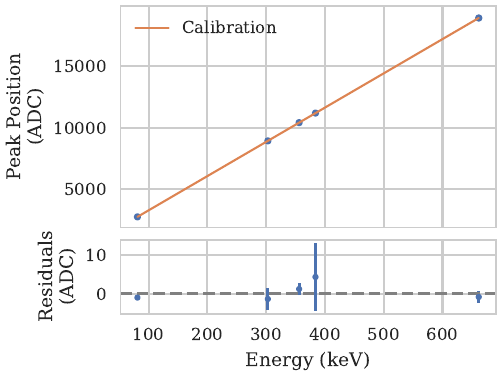}
            \caption{Comparison of the energy calibration obtained from the simultaneous fit to all lines (orange) to the peak positions obtained from independent fits to all groups shown in \cref{fig:photoncalibration} (blue). Figure from \cite{Lebert.2026}.}
            \label{fig:calibration-comparison}
        \end{figure}

        The results of a simultaneous fit to all peaks for a selected pixel and assuming a linear channel-energy relation are shown in \cref{fig:photoncalibration}.
        The reduced $\chi^2$ is in the range of \numlist{1.1;1.3}, which does still indicate some tension, while the residuals do not show systematic deviation.

        We obtain $\sigma \approx \qty{3.1}{\keV}$ for all pixels.
        The gain for pixel 15 was determined to a statistical precision of $\frac{\Delta g}{g} \leq \num{7e-5}$ and the offset to $\frac{\Delta t}{t} \leq \num{4.5e-4}$. All other pixels show similar results.
        The resulting statistical fractional uncertainty on the energy is smaller than \num{1e-4} for energies above $\simeq$\,\qty{150}{\keV}.

        To asses systematic uncertainties, we determined the influence of the backscatter peak shift of the \qty{80}{\keV} line by simulation, varied the choice of background model and binning, and investigated the inclusion of a second-order term in the calibration function (see below). In total, the conservative fractional systematic uncertainty for the pixels indicated in \cref{fig:silicon_detector} is in the range $2\ldots8\times 10^{-4}$ with the exception of a single outlier. The uncertainty contribution on the beta asymmetry parameter $A$ would be smaller than \num{1e-4}. We note that a precision of the Fierz term of $10^{-3}$ requires the fractional uncertainty on the gain to be $\mathcal{O}(10^{-4})$.

        To accommodate for potential non-linear behaviour of the detector system, a second-order term was included in the calibration.
        Fits including this term yield a contribution on the order of $\mathcal{O}\left(\num{e-6}\right)$ for all pixels and with the $\chi^2$ reduced by less than \num{0.02} for most pixels.
        The fractional integral nonlinearity of the ADC was determined to be $\approx 10^{-4}$ \cite{Lebert.2026}.
        The comparison in \cref{fig:calibration-comparison} of the result of the linear calibration obtained from the simultaneous fit to all lines and fits to individual lines without assumptions on the calibration function shows very little tension.

        While we use photons to characterize the system in this work, \citeauthor{Taylor.2025} use X-rays and conversion electrons to calibrate the detector system up to an energy of about \qty{390}{keV}, and \citeauthor{Gonzalez.2026} report on a calibration with conversion electrons up to about \qty{1}{MeV} with electrons with a relative precision of $6.7\times10^{-3}$ using a polynomial.

    \subsection{Signal Rise Time}

        While PERC does not have as stringent requirements on the timing information as Nab (compare \cite{Hayen.2023}), the speed of the signal formation is still important for the planned use case within PERC.
        Only with a quick rise time, the data acquisition can be triggered within $\mathcal{O}(\qty{10}{\ns})$ and coincident measurements be performed with the upstream detector system to measure and potentially veto backscattering events.

        \label{sec:risetime}
        \begin{figure}
            \centering
            \includegraphics[width=\linewidth]{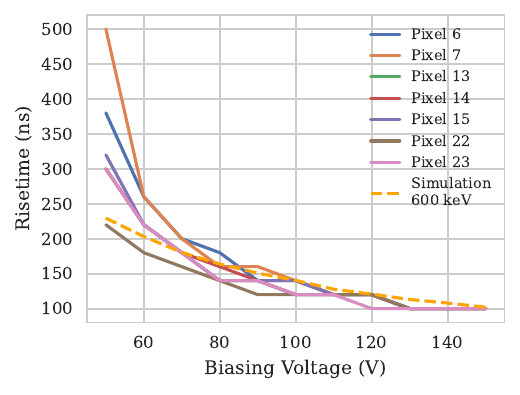}
            \caption{Signal rise time of the detector signal for electrons in the energy range of \qtyrange{410}{510}{\keV} for different pixels.
            The measurement results are compared to results from a Monte Carlo simulation.
            The simulations are in very good agreement with the measurements.
            The spread at low biasing voltages can be explained by differences in depletion voltages across pixels.}
            \label{fig:risetime}
        \end{figure}

        To determine the signal rise time, the \textquote{calibration setup} with an open \isotope{137}{Cs} source was used.
        The signal waveform was sampled at a rate of \qty{50}{\mega\hertz}. 
        The rise time is taken to be the time the signal takes to go from \qtyrange{10}{90}{\percent} of its maximum. 
        To remove the effect of noise and energy dependence, $\approx$\,\num{3000} waveforms within a \qty{100}{\keV} wide energy window have been averaged.
        The measured times for different biasing voltages are shown in \cref{fig:risetime}.
        The decreasing spread in the rise times can be explained by the differing depletion voltages as discussed in \cref{sec:leakage_current}.
        Increasing the biasing voltage to a level well above the depletion voltage, as expected, decreases the signal rise time.
        The measurement was stopped at \qty{150}{\volt} when the guard ring reached the power-supply's maximum current limit, preventing it from maintaining the configured voltage.

        Monte Carlo simulations have been performed using PENELOPE \cite{Penelope.2019}:
        Monoenergetic electrons were targeted at a \qty{2}{\mm} thick silicon slab. 
        The deposited energy per electron at various depths in the silicon was tallied.
        The signal was calculated using the parallel-plate geometry from \cite{Spieler.2012}, assuming a uniform electric field within the detector.
        The doping concentration within the detector can be determined using the graph of the capacitance as a function of the bias voltage, a so-called C\textsuperscript{2}-V-plot \cite{Spieler.2012}.
        The manufacturer provided such a measurement.
        According to the specification of the detector, the concentration should be \qty{1.85e11}{\per\cm\cubed}, but is estimated to be \qty{2.47(4)e10}{\per\cm\cubed} from the capacitance measurements by the supplier. This result is in agreement with similar measurements in~\cite{Hayen.2023}.
        This significant difference also explains the notable deviation of the full-depletion voltage from the specified \qty{300}{\volt}. 
        Using our estimated concentration value in the signal calculation, we obtain the dashed curve in \cref{fig:risetime}.


    \subsection{Electron Detection}

        The energy deposition in silicon by electrons differs from that by photons because electrons deposit energy along their path, while photons mainly interact at a single point.
        Consequently, the peak structure differs slightly.
        For the analysis of monoenergetic electrons from \isotope{207}{Bi} a convolution of the following functions is used \cite{Berger.1969,Steinbauer.1994}:
        \begin{align}
            f_1(u) &= \frac{1}{\sqrt{2\pi\sigma^2}} \exp\left(-\frac{u^2}{2\sigma^2}\right)\\
            f_2(u) &= a\cdot \exp\left(\frac{u}{\tau}\right)\cdot\theta(-u) + b \cdot\theta(-u) + c
        \end{align}
        The first function describes the monoenergetic line, including the Fano and electronic noise.
        The second function consists of an exponential decay to account for the production of Bremsstrahlung, a step function to approximate the effect of electron backscattering, and a constant background. 
        The parameter $u=E-\mu$ is the relative energy to the peak energy $\mu$, and $\theta(u)$ is the Heaviside function.
        
        \label{sec:electrondetection}
        \begin{figure}
            \centering
            \includegraphics[width=\linewidth]{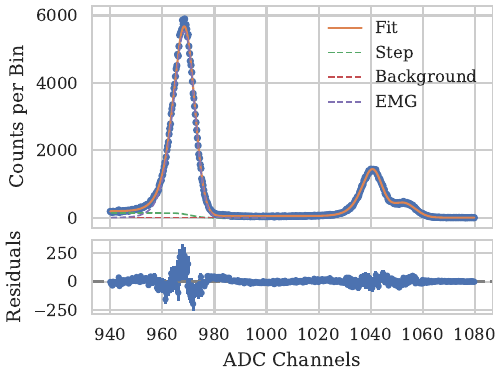}
            \caption{Measured electron response for the K-, L-, and M-conversion electrons at at \qtylist{975.651; 1047.795; 1059.805}{\keV}, together with the fitted response function. EMG stands for \enquote{Exponentially Modified Gaussian}. The residuals indicate that the fit function needs further refinement in the future.}
            \label{fig:bi_975keV}
        \end{figure}

        \Cref{fig:bi_975keV} shows the measured spectrum and the fitted function for the K-, L-, and M-conversion electrons from decay of \isotope{207}{Bi} at \qtylist{975.651; 1047.795;1059.805}{\keV}, respectively \cite{Kondev.2011}.
        The relative positions of the lines were fixed and the energy of the K-line was used as a fit parameter using the calibration obtained with the photopeaks.
        The measurement was carried out at ambient pressure. A simulation taking the geometry of the source and the air into account, shows an energy loss of about \qty{4.3}{\keV}. The measured peak position was \qty{970.1(1)}{\keV} on average across all pixels, in reasonable agreement with our simulation.

        The observed width of the peaks for electrons is slightly higher compared to that for photons and is in the range between \qtylist{3.6;4}{\keV}.
        The increase can at least in part be attributed to the window on the source and the air, which both influence the charged particles but are irrelevant for photons. Our simulations show an expected increase in the observed width of \qty{0.3}{\keV}.

        Only less than \qty{0.5}{\percent} of the electrons at this energy, which hit the detector with a pitch angle of \ang{90}, can actually penetrate the detector.
        The average energy loss of those electrons is \qty{1.6(2)}{\keV}, which leads to a distortion of the spectrum by less than \qty{10}{\eV}, which is a negligible shift.

        In PERC, we will use open electron conversion sources deposited on very thin foils and located close to the central uniform field to calibrate the detector. This ensures almost unperturbed electron spectra and an angular distribution of the electron impact on the detector which is nearly identical to that of electrons from neutron decay.

    \subsection{Charge Sharing}
        \label{sec:chargesharing}
        Charge sharing occurs when the charge cloud generated by a single electron spreads across multiple adjacent pixels before it is fully collected. 
        As a result, the deposited energy is split between neighbouring pixels.
        This effect arises from diffusion and drift of charge carriers in the detector bulk and occurs near pixel boundaries. 
        The boundary between pixels is \qty{100}{\um} wide, but only on the anode side.
        This means that the electric field lines split, thus, all charges are transported to one pixel or the other \cite{Hayen.2023}.
        Therefore, no energy is lost due to the pixel structure.

        To investigate this effect, the \textquote{dead layer setup} was used with a \isotope{207}{Bi} source instead of the alpha source.
        The source has an active diameter of \qty{5}{\mm}, which is similar in size to the pixels. Being placed at a distance from the detector of \qty{4}{cm}, it illuminates the entire pixel.

        Two effects influence the amount of charge sharing: the lateral diffusion of the charge cloud within the detector, and the drift of the primary particle.
        Electrons with an energy of \qty{1}{\MeV} have a (CSDA) range of around \qty{2.3}{\mm} in silicon. 
        This is also the maximum drift length of those electrons.
        
        The lateral diffusion of the charge cloud $\sigma$ can be calculated by \cite{Spieler.2012}
        \begin{equation}
            \sigma = \sqrt{2Dt}\quad \mbox{with}  \quad D = \frac{k_B T}{e}\mu.
        \end{equation}
        The second equation is the Einstein–Smoluchowski relation, which links the mobility of the charge carrier $\mu$ to the diffusion constant $D$, with Boltzman's constant $k_B$,  the electric charge of the electron $e$, and the temperature $T$.
        Assuming a collection time $t$ of \qty{250}{\ns}, we obtain a spread of around \qty{40}{\um}, which is negligible in comparison to the drift of the primary particles.

        \begin{figure}
            \centering
            \includegraphics[width=\linewidth]{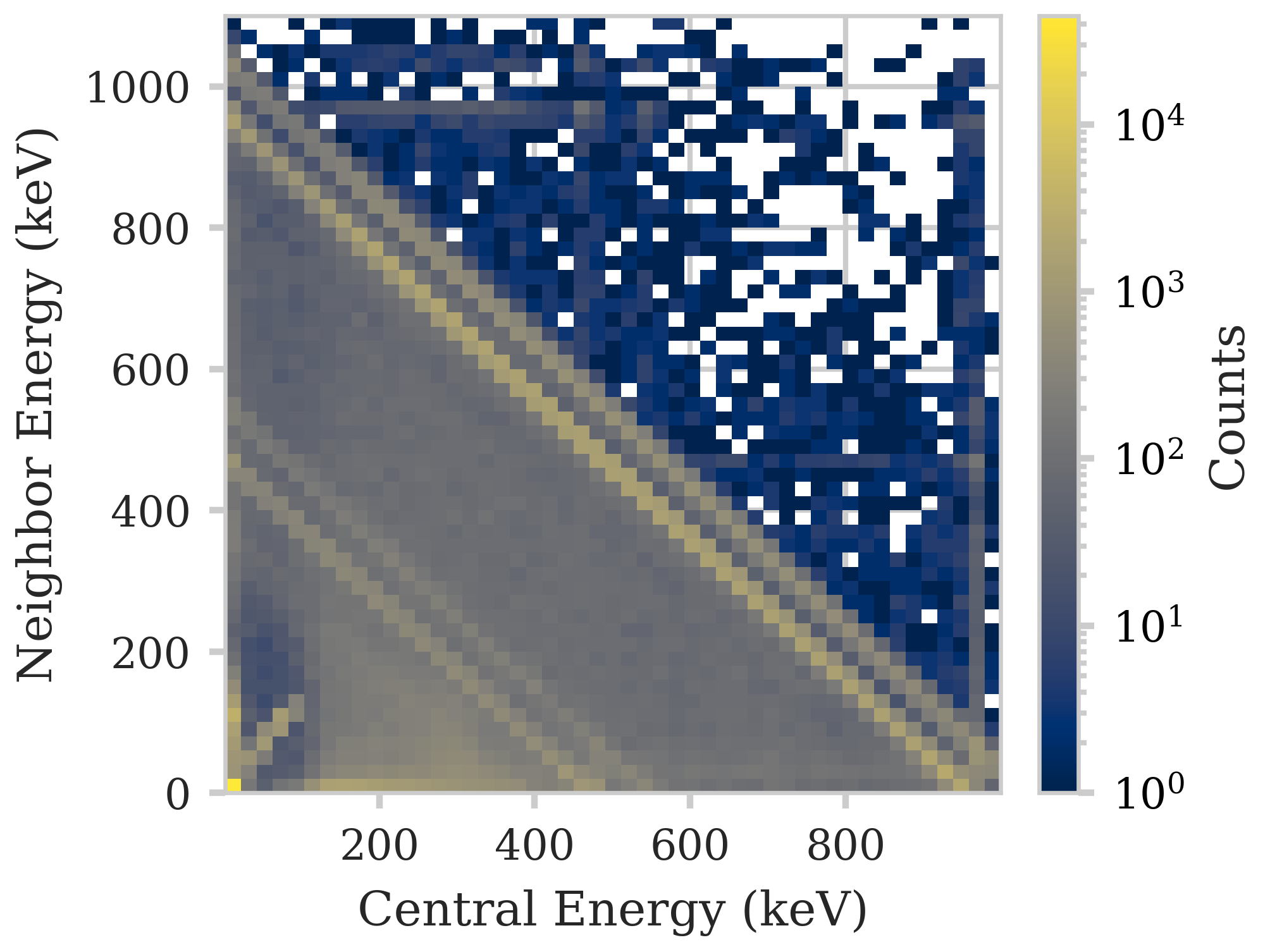}
            \caption{Energy distribution in neighbouring pixels 14 and 15 for events from a \isotope{207}{Bi} source. Lines with negative slope indicate charge-sharing at the energies of the conversion electrons. Vertical and horizontal features indicate random coincidences. Figure from \cite{Lebert.2026}.}
            \label{fig:charge-sharing}
        \end{figure}

        To measure the effect of charge sharing, all neighbouring pixels were read out simultaneously with every trigger event of the central pixel, see \cref{fig:silicon_detector}. \Cref{fig:charge-sharing} exemplarily shows the distribution of energy between the central pixel~14 and its neighbour pixel~15. Charge-sharing events of the conversion electrons are clearly identifiable by the diagonals with negative slope. Vertical and horizontal features indicate random coincidences. Events without charge-sharing are on the horizontal axis. Since we require the central pixel to trigger, events near the vertical axis are all coincidences with noise in the central pixel. The diagonal at low energies are coincidences of the many electrons and X-rays in range \qtyrange{70}{90}{keV} \cite{Be.2010}, and noise triggers around \qty{50}{keV}.

        To investigate effects of charge-sharing on the measured energy, we now consider the peak in the spectrum at \qty{975}{keV}, where we expect the largest impact in the relevant energy range.
        We select events for which the measured energy sum is in the range \qtyrange{900}{1000}{keV}, which corresponds to a diagonal region around the higher energy lines in \cref{fig:charge-sharing}.
        To select only events that share their energy between two pixels, we require the energy in each pixel to be smaller than \qty{895}{\keV}. This corresponds to vertical and horizontal limits in \cref{fig:charge-sharing}. This way we require at least \qty{60}{\keV} to be shared for the conversion line at \qty{975}{keV}.
        This ensures that only random coincidence events could distort the resulting spectrum, which is neglected here.
        For three-pixel events, the energy in every pixel is required to be in the range \qtylist{25;895}{\keV}.
        The higher lower limit is necessary to remove two-pixel events by ensuring that the shared energy among the pixels is sufficient to exceed the noise level.

\begin{figure}
            \centering
            \includegraphics[width=\linewidth]{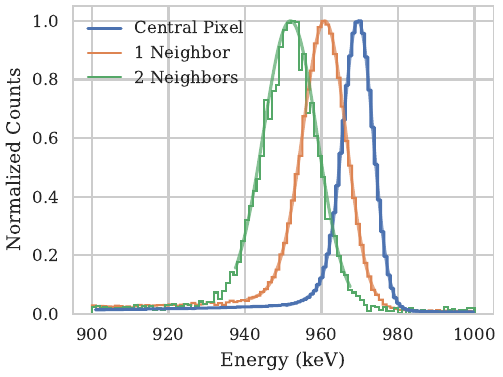}
            \caption{Resulting energy peak of the \qty{975}{\keV} conversion electron peak of \isotope{207}{Bi} for charge shared events.
            The central pixel peak comprises all events of the central pixel in this energy region.
            The other two peaks are obtained by summing the energy of time-correlated events involving two or three pixels.
            Energy cuts have been applied to remove noise.
            All peaks are normalised to their maximum to facilitate comparison.}
            \label{fig:charge_share_shift}
        \end{figure}

        In \cref{fig:charge_share_shift}, the resulting spectra are shown.
        We observe a significant shift in the peak position and an increase in the width of the peak compared to the non-coincident read-out.
        The change in width is due to the fact that now several pixels with their inherent resolutions are used in the data acquisition, which leads to a factor of about \num[parse-numbers=false]{\sqrt{2}} and \num[parse-numbers=false]{\sqrt{3}} for the peaks of two and three pixels, respectively (assuming all resolutions are the same).
        A likely explanation of the observed energy shift with respect to the single pixel peak of about \qty{6.8}{keV} and \qty{14.8}{keV} are parasitic capacitances between channels, which need only be on the order of a few femtofarads to have a significant effect \cite{Pullia.2011}. This is consistent with the observation that the shift about doubles for three-pixel events in comparison to two-pixel events. Simulations exclude an effect larger than \qty{0.1}{keV} from the larger angles of incidence on neighbouring pixels. No change of the energy shift and width of the two-pixel events with the ratio of energy is observed. An equivalent analysis using \qty{480}{keV} conversion electrons yields similar relative effects.

        Ref.~\cite{Taylor.2025} investigates the integral contribution to cross-talk by analysing the recorded waveforms of a central pixel and its six neighbours. It finds a negative coupling due to the mutual capacitance of the pixels and contributions from the preamplifier in the range of \qtyrange{0.4}{0.9}{\percent} which is similar in sign and magnitude to our findings.

        With the energy cuts applied, the observed fraction of charge shared events between two neighbouring pixels is in the range \qtyrange{4.8}{4.9}{\percent}. 
        The fraction of events shared between three pixels is around \qty{0.3}{\percent}.
        In total, nearly \qty{31}{\percent} of \qty{975}{\keV} events are shared with neighbouring pixels.
        
        To estimate the expected fraction of charge-sharing events, we simulate the radial energy distribution of electrons impinging on the silicon with the simplifying assumption of perpendicular impact.
        From it we calculate the percentage of particles that deposit more than a threshold energy $E_\mathrm{cs}$ outside the inner pixel for uniformly distributed impact positions across the pixel.
        We choose $E_\mathrm{cs} = \qty{60}{\keV}$, matching the energy cut applied to the measurements.
        We obtain a fraction of \qty{28}{\percent} of events to be charge-sharing events for an electron energy of \qty{1}{\MeV}.
        This estimate is slightly lower than the measurement result above. The more tilted distribution of angles of incidence should increase that value. Charge-sharing and cross-talk at the percent level are highly significant effects for precision beta spectroscopy and require further studies.

    \subsection{Dead Layer}
        \label{sec:deadlayer}
        On the entrance window of the detector is an insensitive layer of silicon dioxide, and the p\textsuperscript{+}-doped layer of the pn-junction.
        This part of the detector is typically referred to as a \emph{dead layer}, because not all energy deposited here is collected.
        A simple model of charge collection efficiency as a function of the depth in the detector is a step function: no charge collection within the dead layer, and full charge collection beyond it.
        This model is incomplete \cite{Hayen.2023, Gugiatti.2020}, but for a dead layer of only \qty{100}{\nm}, which we expect, it is sufficient for a first characterisation.

        The source includes \isotope{233}{Pu} with a line at \qty{5.155}{\MeV}, \isotope{241}{Am} at \qty{5.486}{\MeV}, and \isotope{244}{Cm} at \qty{5.805}{\MeV}.
        Three measurements were performed: one with the source positioned centrally below the detector, one at a distance corresponding to a \ang{60} pitch angle of incident \textalpha-particles, and one with the source centrally below the detector but inclined by \ang{60}.
        By changing the incident angle $\delta$ of the incoming particles, the dead layer is effectively increased by a factor of $\cos^{-1}\delta$ \cite{Knoll.2010}.
        With the reasonable assumption that the atomic scattering and recombination contributions to the pulse height defect do not depend on the incident angle, we determine the thickness of the dead layer of the detector and the source from the energy shifts of the alpha lines.

        We obtain a thickness of \qty{123(6)}{\nm} silicon equivalent averaging results from six peaks where the error is determined from the systematic variation of the results for the different lines.
        This is more than the specified \qty{100}{\nm}. Potential issues in the analysis could be channelling, see \cite{Beck.2024}. We note that Measurements in Ref.~\cite{SalasBacci.2014} with protons on a \qty{1}{\milli\meter} thick detector indicate a thickness in the range \qtyrange{70}{110}{\nano\meter}. Ref.~\cite{Taylor.2025} extends the analysis to a \enquote{soft} and \enquote{hard} contribution of the dead layer with about equal thicknesses in the range \qtyrange{55}{60}{\nano\meter}.

        However, the dead layer results in an energy shift of only \qty{30}{\eV} for a \qty{50}{\keV} electron or \qty{0.6}{\text{\textperthousand}}. It is hence a minor effect for electron measurements.

\section{Summary}

    In this work, we present a characterisation of some pixels of a large area pixelated silicon PIN detector at room temperature using commercial electronics to be used in the PERC experiment for neutron $\beta$-spectroscopy.
    It was originally designed for the Nab experiment \cite{Gonzalez.2026}.

    Using photons, we demonstrate that the detector is highly linear, with a calibration uncertainty below \qty{100}{\eV}. 
    The detector system has a good energy resolution of approximately \qty{3.1}{\keV} (FWHM of \qty{7.5}{\keV}) for photons and \qty{4}{\keV} (FWHM of \qty{9.4}{\keV}) for electrons with the detector at room temperature.
    With the current setup, charge sharing for electron events leads to a loss in detected energy. It may be attributed to small parasitic capacitance between channels. This might be problematic with regard to the envisaged measurement precision for the Fierz interference term and requires further studies.

\section*{Acknowledgements}

    This work is supported by the Priority Program SPP~1491 of the German Research Foundation (DFG), and the Austrian Science Fund (FWF), the FRM~II, and the DFG cluster of excellence ORIGINS under Germany’s Excellence Strategy – EXC-2094 – 390783311.

    The PERC project is a collaborative effort with members of the Technical University of Munich, the TU~Vienna, the University of Heidelberg, the Johannes Gutenberg University Mainz, the Institut Laue-Langevin, the Paul-Scherrer-Institut, and the FRM~II.

    We thank Ulrich Schmidt, University of Heidelberg, for helpful discussions.

\bibliographystyle{elsarticle-num-names-max} 
\bibliography{references.bib}

\end{document}